\documentstyle[11pt,aaspp4]{article}

\newbox\grsign \setbox\grsign=\hbox{$>$} 
\newdimen\grdimen \grdimen=\ht\grsign
\newbox\laxbox \newbox\gaxbox
\setbox\gaxbox=\hbox{\raise.5ex\hbox{$>$}\llap
     {\lower.5ex\hbox{$\sim$}}}\ht1=\grdimen\dp1=0pt
\setbox\laxbox=\hbox{\raise.5ex\hbox{$<$}\llap
     {\lower.5ex\hbox{$\sim$}}}\ht2=\grdimen\dp2=0pt

\begin{document}

\title{
High Resolution Spectroscopy of the X-ray Photoionized Wind in
Cygnus X-3 with the {\it Chandra} High Energy Transmission Grating
Spectrometer
}

\author{
Frits Paerels\altaffilmark{1,2}, 
Jean Cottam\altaffilmark{1},
Masao Sako\altaffilmark{1},
Duane A. Liedahl\altaffilmark{3}, 
A. C. Brinkman\altaffilmark{2},
R. L. J. van der Meer\altaffilmark{2},
J. S. Kaastra\altaffilmark{2}, and
P. Predehl\altaffilmark{4}
}
\altaffiltext{1}{Columbia Astrophysics Laboratory,
Columbia University, 538 W. 120th St., New York, NY 10027, USA} 

\altaffiltext{2}{SRON Laboratory for Space Research,
Sorbonnelaan 2, 3584 CA Utrecht, the Netherlands}

\altaffiltext{3}{Department of Physics,
Lawrence Livermore National Laboratory,
P.O. Box 808, L-41, Livermore, CA 94550, USA}

\altaffiltext{4}{Max Planck Institut f\"ur Extraterrestrische
Physik, Postfach 1503, D-85740 Garching, Germany}

\begin{abstract}

We present a preliminary analysis of the 1--10 keV spectrum of the
massive X-ray binary Cyg X-3, obtained with the High Energy
Transmission Grating Spectrometer on the {\it Chandra X-ray
Observatory}. The source reveals a richly detailed discrete emission
spectrum, with clear signatures of photoionization-driven excitation. 
Among
the spectroscopic novelties in the data are the first astrophysical
detections of a number of He-like 'triplets' (Si, S, Ar)
with emission line
ratios characteristic of photoionization equilibrium, fully resolved
narrow radiative recombination continua of Mg, Si, and S, 
the presence of
the H-like Fe Balmer series, and a clear detection of a $\sim 800$ km
s$^{-1}$ large scale velocity field, as well as a $\sim
1500$ km s$^{-1}$
FWHM Doppler broadening
in the source. We briefly touch on
the implications of these findings for the structure of the Wolf-Rayet
wind.


\end{abstract}

\section{Introduction}

In a previous paper (Liedahl \& Paerels 1996, 'LP96') we presented an
interpretation of the discrete spectrum of Cyg X-3 as observed with
the Solid State Imaging Spectrometers on {\it ASCA} (cf. Kitamoto
et al. 1994; Kawashima \& Kitamoto 1996). 
We found clear spectroscopic evidence that the discrete emission is
excited by recombination in a tenuous X-ray
photoionized medium, presumably
the stellar wind from the Wolf-Rayet companion star (van Kerkwijk et  
al. 1992). 
Specifically, the {\it ASCA} spectrum revealed a narrow
radiative recombination continuum (RRC) from H-like S, unblended
with any other transitions. On closer inspection, RRC features due to
H-like Mg  and Si were also found to be present in the data, although
severely blended with emission lines. These narrow continua are an
unambiguous indicator of excitation by recombination in X-ray
photoionized gas, and their relative narrowness is a direct
consequence of the fact that a highly ionized photoionized 
plasma is generally much cooler than a collisionally ionized plasma of
comparable mean ionization (LP96, Liedahl 1999 and references
therein).

With the high spectral resolution of the {\it Chandra} High Energy
Transmission Grating Spectrometer, we now have the capability to fully
resolve the discrete spectrum. 
Apart from offering a unique way to determine the structure of the
wind of a massive star, study of the spectrum may yield
other significant benefits. Cyg X-3 shows a bright, purely
photoionization driven spectrum, and, as such, may provide a template
for the study of the spectra of more complex accretion-driven
sources, such as AGN. The analysis will also allow us to verify
explicitly 
the predictions for the structure of X-ray photoionized nebulae
derived from widely applied X-ray photoionization codes.

\section{Data Reduction}

A description of the High Energy Transmission Grating Spectrometer
(HETGS) may be found in Markert et al. (1994).
Cyg X-3 was observed on October 20, 1999, for a total of 14.6 ksec
exposure time, starting at 01:11:38 UT.
The observation covered approximate binary phases $-0.31$ to 
$+0.53$, which
means that about half of the exposure in our observation occurs in the
broad minimum in the lightcurve at orbital phase zero.
Aspect-corrected data from the standard CXC pipeline 
(processing date October 30, 1999) was post-processed 
using dedicated procedures written at Columbia. We used 
({\it ASCA}-)grade 
0,2,3,4 events, 
a spatial filter 30
ACIS pixels wide was applied to both the High Energy 
Grating (HEG) and Medium Energy Grating (MEG)
spectra, and
the resulting events were plotted in a dispersion--CCD pulse height
diagram, in which the spectral orders are neatly separated. 

A second filter was applied in this dispersion--pulse height diagram.
The filter consisted of a narrow mask centered on each of the spectral
orders separately. The mask size and shape were optimized
interactively.
The
residual background in the extracted spectra is of order 0.5
counts/spectral bin of 0.005 \AA\ or 
less. 
The current state of the calibration does not provide us with the
effective area associated with our joint spatial/pulse height 
filters to better than 25\% accuracy, hence we have chosen not to
flux-calibrate the spectrum at this time.
An additional correction to the flux in the chosen aperture
due to the (energy
dependent) scattering of photons by interstellar dust has not yet
been determined either.

In the resulting order-separated count spectra, we located the zero
order and we determined its centroid position to find the zero of the
wavelength scale. We then converted pixel number to wavelength based
on the geometry of the HETGS. In this procedure, we used ACIS/S chip
positions that
were determined after launch from an analysis of the dispersion
angles in the HETGS spectrum of Capella (Huenemoerder et al. 2000). 
This preliminary wavelength
scale appears to be accurate to approximately 2 m\AA. 
The spectral resolution was
determined from a study of narrow, unblended
emission lines in the spectrum of Capella.
It is approximately constant accross the entire HETGS band, and amounts
to approximately 0.012 \AA\  (0.023 \AA) FWHM for the HEG (MEG) (Dewey
2000).
The resolution in the Cyg X-3 spectrum can be checked
self-consistently by analyzing the width of the zero order image.
Unfortunately, the zero order image is affected by pileup. 
However, enough
events arrive during the 41 ms CCD frame transfer, forming a
streak in the image, 
that we can construct an unbiased 1D zero-order
distribution from them. The width of this distribution is consistent
with the widths of narrow lines in the spectrum of Capella, which
indicates that the resolution in the Cyg X-3 spectrum is not affected
by systematic effects ({\it e.g.},
incorrect aspect solution, defocusing).

\section{X-ray Photoionization in Cyg X-3}

Figure 1 shows the HEG and MEG first order spectra; the higher order
spectra are unfortunately very weak, and we will not discuss them
here.
We show the spectra as a function of wavelength, because this is the
most natural unit for a diffractive spectrometer: the instruments have
approximately constant wavelength resolution.
The spectra have been smoothed with a 3-pixel boxcar average to
bring out coherent features 
We have indicated the
positions of expected strong H-- and He-like discrete features. A
cursory examination of the spectrum strikingly confirms the
photoionization-driven origin of the discrete emission. 

We detect the spectra of the H-like species of all abundant 
elements from Mg through Fe. In Si and S, we detect well-resolved 
narrow radiative recombination continua. This is illustrated in Figure
2, which shows the 3.0--7.0 \AA\ band on an enlarged scale. The 
Si XIV and
S XVI continua are readily apparent.
The width of these features is a
direct measure of the electron temperature in the recombining plasma,
and a simple eyeball fit to the shapes indicates $kT_e \sim$ 50 eV,
which is roughly in agreement with the result of model calculations
for optically thin X-ray photoionized nebulae (Kallman \& McCray
1982). A more detailed, fully quantitative analysis of the spectrum
will be required to see whether we can also detect the 
expected temperature gradient in the source (more highly ionized zones
are also expected to be hotter).
In the Si XIV and S XVI spectra we estimate the ratio between 
the total photon flux in the RRC to that in Ly$\alpha$ to be about 
0.8 and 0.7, respectively; here, we assume $kT_e =$ 50 eV, and we have
made an approximate correction for the differences in effective area
at the various features. These measured ratios
are in reasonable agreement with the expected ratio of 
$0.73 (kT_e/20\ {\rm eV})^{+0.17}$ (LP96), which indicates that the
H-like spectra are consistent with pure recombination in optically
thin gas.

The positions of the lowest members of the Fe XXVI Balmer series are
indicated in Figure 1 (the fine structure splitting
of these transitions is
appreciable in H-like Fe, as is evident from the plot). The
relative brightness of the Balmer spectrum is yet another indication of
recombination excitation.
There is evidence for line emission at the position of H$\beta$, and
possibly at H$\gamma$ and H$\delta$; the spectrum is unfortunately too
heavily absorbed to permit a detection of H$\alpha$ ($\lambda\lambda
9.52,9.74$ \AA). Unfortunately, the long-wavelength member of the 
H$\beta$ 'doublet'
($\lambda \approx 7.17$ \AA) almost precisely coincides
with the expected position of Al XIII Ly$\alpha$, which precludes 
a simple and neat direct detection of Al (the first detection of an
odd-$Z$ triple-$\alpha$ element in non-solar
X-ray astronomy). Any limit on the
Al/Si abundance ratio thus becomes dependent on an understanding of the
intensity of the Fe XXVI spectrum.

As for the He-like species, we detect the $n=2-1$ complexes, consisting
of the forbidden ('$f$'), intercombination ('$i$'), and resonance 
('$r$') transitions, in Si
XIII, S XV, Ar XVII, Ca XIX, and Fe XXV (as well as the corresponding
RRC in Si, S, and possibly Ar). The line complexes appear resolved
into blended resonance plus intercombination lines, and
the forbidden line (see Figures 1 and 2), up to Ar XVII.

In an optically thin, low density,  
purely photoionization-driven plasma, one expects the intensity ratio
$f/(r+i) \approx 1$
for the mid-$Z$ elements, very different from the pattern in the more
familiar collisional equilibrium case, where the resonance transition
is relatively much brighter ({\it e.g.}, Gabriel \& Jordan
1969; Pradhan 1982; Liedahl 1999). We use the ratio $f/(r+i)$ rather
than the conventional $G \equiv (i + f)/r$ and $R \equiv f/i$,
because the intercombination and resonance lines are unfortunately 
blended by significant Doppler broadening in the source
(see Section 4).
Theoretically, in a photoionized plasma 
$f/(r+i)$ is approximately equal to 1.3, 1.0, 0.83, for
Si XIII, S XV, and Ar XVII, respectively, and 
depends only weakly on electron temperature (LP96, Liedahl
2000). The measured ratios $f/(r+i)$, derived by fitting three
Gaussians with common wavelength offset and broadening at the expected
positions of $f, i$, and $r$, are approximately 1.1, 0.8, and 1.1
with the
HEG, for Si, S, and Ar, respectively; the corresponding ratios for 
the MEG are 1.3, 1.0, and 0.8. Since most of the lines contain at least
100 photons, the statistical error on the ratios is generally less
than 15\%. These measurements include a model for the Si XIII RRC in
the S XV triplet (assuming $kT_e = 50$ eV), and Mg XII Ly$\gamma$
emission in the Si XIII triplet.

The He-like line ratios are probably
affected by systematic features in the efficiency of the
spectrometer. The S XV triplet is superimposed on the Si XIII
RRC, the Si XIII triplet straddles the Si K edge in the CCD
efficiency, and the Ar XVII triplet straddles the Au M$_{\rm IV}$ 
and Ir M$_{\rm I}$ edges. 
Corrections for these effects will have to be carefully
evaluated. Nevertheless, the raw ratios $f/(r+i)$ for the Si and Ar
triplets are already of the right magnitude for pure recombination. 
Our provisional conclusion is that the 
He-like spectra are, very roughly, consistent with pure
recombination in optically thin gas.

Just as in a collisional plasma, the relative strengths of the
forbidden and intercombination lines are sensitive to density (Liedahl
1999; Porquet \& Dubau 2000), due to collisional transfer between the
upper levels of $f$ and $i$ at high density. As mentioned above,
there are some systematic
uncertainties in the measured line ratios, and we defer a discussion
of possible constraints on the density in the wind to a future paper.

The detection of fluorescent Fe emission
is a surprise, because virtually
no fluorescence
was seen at the time of the {\it ASCA} observation
(Kitamoto et al. 1994). The apparent
centroid wavelength of the fluorescent line is $1.939$ \AA\
(photon energy 6394 eV), with a
formal error of less than $10^{-3}$ \AA\ (3 eV). 
The width of the line is 
$0.022$ \AA\ FWHM, with a formal uncertainty of less than 5\%.
This is wider than would be expected from the 
velocity broadening to be discussed in the next section, and may
be an indication that a range of ionization stages contributes to the
fluorescent emission. If we assume the same velocity broadening for
the Fe K$\alpha$ feature as for the high-ionization lines (which may 
not necessarily be correct if the low-- and high--ionization lines
originate in different parts of the stellar wind), we find that 
Fe K$\alpha$ has an intrinsic width (expressed as the FWHM of a
Gaussian distribution) of 0.018 \AA\ (corresponding to $\Delta E 
\approx 60$ eV). The fine structure split between K$\alpha_1$ and 
K$\alpha_2$ contributes slightly 
to this width ($\Delta\lambda \approx 0.004$
\AA), but the measured width covers the full range of 
K$\alpha$ wavelengths for charge states between fully neutral and 
Ne-like (Decaux et al. 1995).

\section{Bulk Velocity Fields}

We find that all emission features are significantly
broadened and redshifted. The lines and radiative recombination continua
are resolved by both the HEG and the MEG. The line widths for
H-like Mg, Si, S, Ar, Ca, and Fe Ly$\alpha$ were measured by
fitting a simple Gaussian profile. Other than the negligibly small
fine structure split ($\Delta\lambda \sim 0.005$ \AA), these lines are
clean and unblended. 
The resulting widths do not seem to exhibit a strong dependence on
phase. Assuming that the
spectrometer profile is well represented by a Gaussian of width
0.012 \AA\ (0.023 \AA) FWHM for the HEG (MEG), we find that the
broadening of the lines is roughly consistent with a Gaussian velocity
distribution, of width $\Delta v \sim 1500$ km s$^{-1}$ FWHM. 
The scatter
is too large to permit a meaningful test for any dependence of the
velocity broadening on ionization parameter. 
Note that no such broadening was
seen in the spectrum of Capella. 

We also measured the radial velocities for the Ly$\alpha$ lines,
assuming the dispersion relation obtained from an analysis of the
spectrum of Capella. Wavelengths were calculated from the level
energies given by Johnson \& Soff (1985); these should be accurate to
a few parts in $10^6$. There is a clear systematic redshift to all the
emission lines and RRCs, in both the positive and negative spectral
orders and in both grating spectra. This is shown in Figure 3, where
we have segregated dim and bright state data, but have averaged
positive and negative spectral orders, and HEG and MEG spectral data. 
Also shown are the best fitting uniform
velocity offsets. These fits were
forced to yield zero wavelength shift at zero wavelength. 
The average redshift for the dim state is $\sim 800$ km s$^{-1}$,
and for the bright state is $\sim 750$ km s$^{-1}$.
We thus
find a net redshift much smaller than the observed velocity 
spread, and essentially no
dependence of the centroid velocity on the binary phase.
We should point out that our preliminary analysis,
based on fitting simple Gaussians, is admittedly crude,
and may have biased the true nature of the velocity field somewhat.
We also note, with caution, that Doppler shifts due to a single, uniform
velocity do not appear to be a very good description of the data: the
longest wavelength lines appear to be offset at a significantly
larger than average radial velocity. A detailed analysis, taking into
account the actual lineshape, will be
required to confirm or refute the possibility that these offsets
represent the expected systematic correlation of average
wind velocity and ionization parameter.

\section{Discussion}

The HETGS spectrum of Cyg X-3 has revealed a rich discrete spectrum,
the properties of which are consistent with pure recombination
excitation in cool, optically thin, low density X-ray photoionized gas
in equilibrium. We fully resolve the narrow RRCs
for the first time, 
and estimate an average electron temperature in the
photoionized region of $kT_e \sim 50$ eV, consistent with global
photoionization calculations.

We detect a net redshift in the emission lines of $v \sim 750-800$
km s$^{-1}$, essentially independent of binary phase, 
and a distribution in velocity with a FWHM of $\sim 1500$ km s$^{-1}$.
If the wind were photoionized throughout, we would expect to
see roughly equal amounts of
blue-- and redshifted material, so evidently we are viewing
an ionized region that is not symmetric with respect to the source of
the wind, as expected if only the part of the wind in the
vicinity of the X-ray continuum source is ionized. 
However, in the simplest wind models, one would then 
expect to see a strong
dependence of the centroid velocity on binary phase, alternating
between red-- and blueshifts, and this is decidedly not the case in
our data. 
The implications of this finding for the flow pattern and
distribution of material in the wind will be explored in a future paper.

Finally, the Fe K$\alpha$ fluorescent feature, which 
probes a more neutral phase of the wind, has never been seen before in
Cyg X-3.
Unfortunately, the exact range of ionization can not
be separated uniquely from systematic Doppler shifts through a
measurement of the wavelengths of the K$\alpha$ spectra, because the
feature, while clearly broadened, is not separated into its component
ionization stages. Still,
the width of the feature (the net effect of the
velocity field and the existence of a range of charge states)
and its intensity
will impose strong constraints on the global properties 
of the wind.

\noindent
Acknowledgements.
\newline
\noindent
We wish to express our gratitude to Dan Dewey and Marten van Kerkwijk,
for discussions and a careful reading of the manuscript, and to the
referee, Randall Smith, for a thorough review.
JC acknowledges support from NASA under a GRSP fellowship.
MS's contribution was supported by 
NASA under Long Term Space Astrophysics
grant no. NAG 5-3541. FP was supported under NASA Contract no. NAS
5-31429.
DL acknowledges support from NASA under Long Term Space Astrophysics
Grant no. S-92654-F. Work at LLNL was performed under the auspices of
the US Department of Energy, Contract mo. W-7405-Eng-48.


\newpage

Figure Captions:

\noindent
Fig.1---The 1--10 \AA\ spectrum of Cyg X-3 as observed with the HEG
(upper panel), and the MEG (lower panel), binned in 0.005 \AA\ bins.
The positive and negative first orders have been added, and the
spectra have been smoothed with a 3 pixel boxcar filter. Labels
indicate the positions of various discrete spectral features.
'He$\alpha$' is the inelegant label for the resonance, 
intercombination, and forbidden lines in the He-like ions, plotted at
the average wavelength for the complex.
High-ionization features of interest 
that were not detected have been labeled in
brackets. Horizontal bars indicate the nominal
positions of the gaps between
the ACIS chips; the dithering of the spacecraft will broaden the gaps
and soften their edges.

\noindent
Fig.2---The 3.0--7.0 \AA\ region of the spectrum enlarged; we show
the raw count rates, binned by two 0.005 \AA\ bins.
The most important transitions have been labeled; dashed lines mark
the expected positions of Si and S recombination edges. These markers
have been redshifted by 800 km s$^{-1}$.
The horizontal bar
near 4.5 \AA\ in the HEG spectrum marks the nominal position of the
gap between chips S2 and S3 in ACIS.
The solid
line in the MEG spectrum is a crude empirical fit to the continuum, with
Si XIII, Si XIV, and S XVI narrow radiative recombination continua
added. The electron temperature was set to 50 eV, and the continua
were convolved with a 1500 km s$^{-1}$ FWHM velocity field, to match
the broadening observed in the emission lines.

\noindent
Fig.3--Measured wavelength shift for selected Ly$\alpha$ features.
Filled symbols refer to the 'dim' state data, open symbols to the
'bright' state data. The velocities as measured with the HEG and the
MEG have been averaged; velocities in
positive and negative spectral orders were
averaged. Error bars indicate the size of the rms variation between
these various measurements. In cases where only one or two velocities
were measurable due to low signal-to-noise, we instead indicate 
the estimated
statistical error on these measurements. The solid lines are the
weighted least squares Doppler velocities for both the dim and the
bright states.

\vfill\eject

\centerline{\null}
\vskip7.5truein
\includegraphics{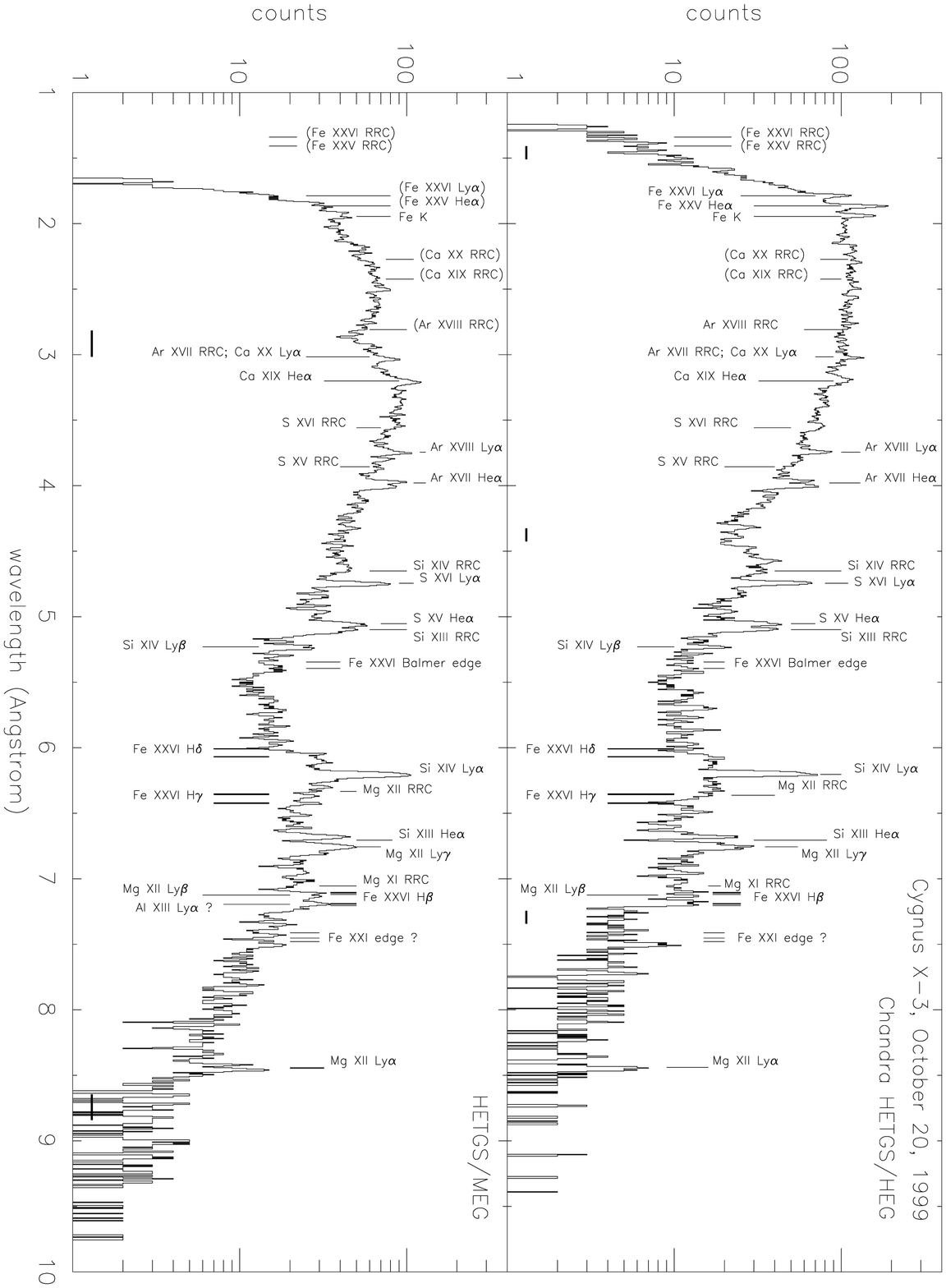}

\vfill\eject

\centerline{\null}
\vskip7.5truein
\includegraphics{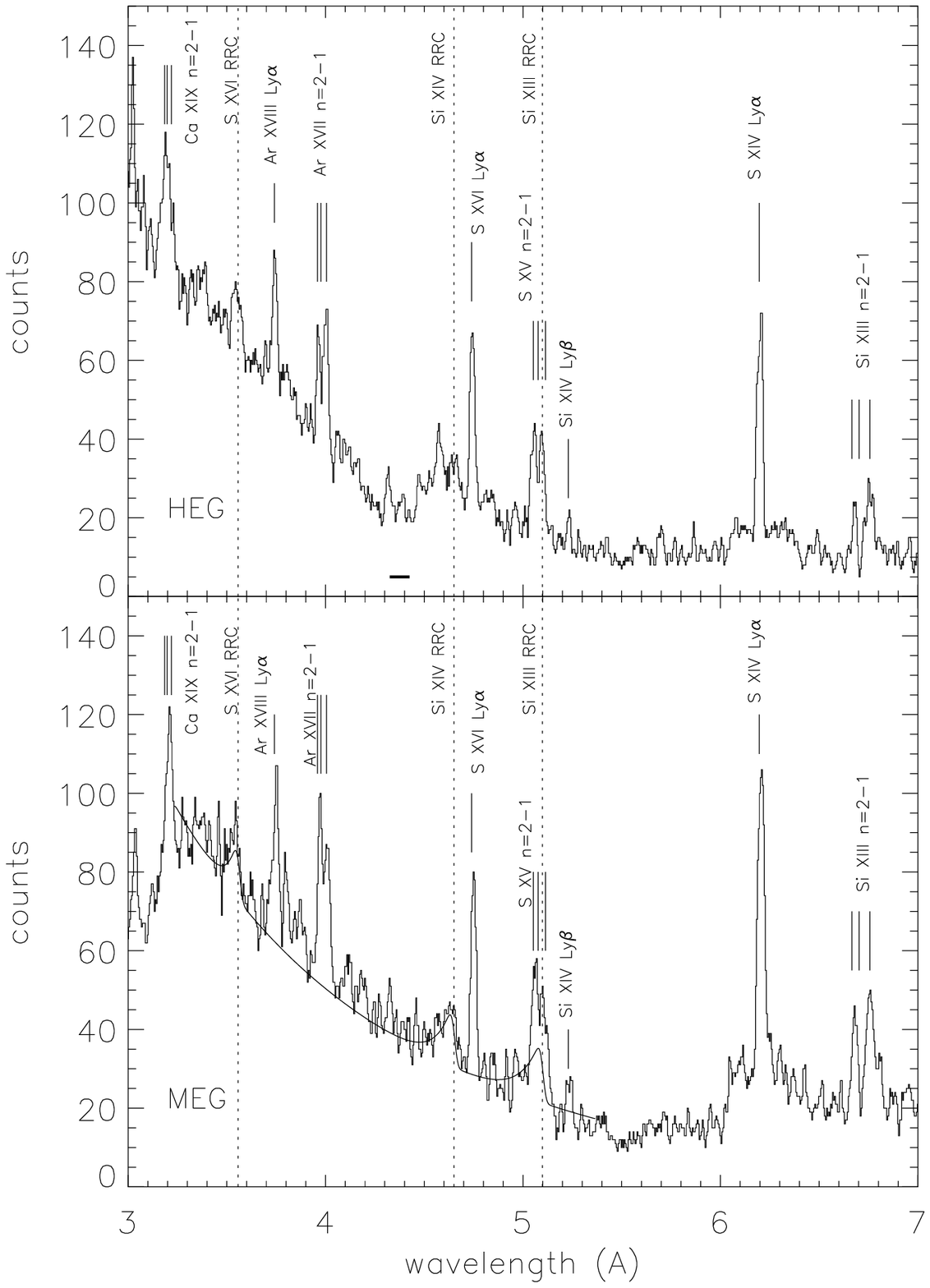}

\vfill\eject

\centerline{\null}
\vskip7.5truein
\includegraphics{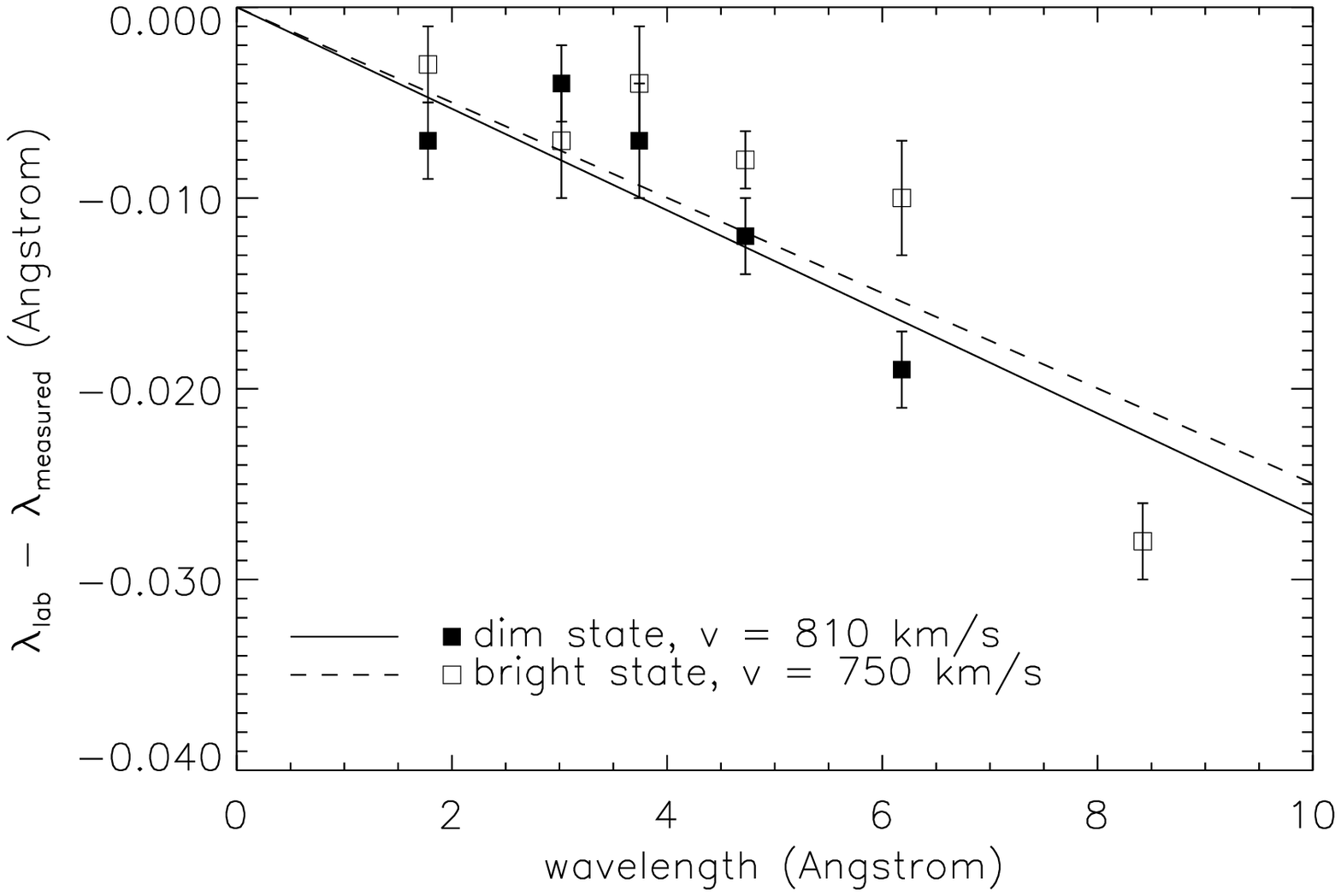}

\end{document}